\input amstex.tex
\documentstyle{amsppt}
\magnification\magstep1
\NoBlackBoxes
\topmatter
\title Gerbes and Quantum Field Theory
\endtitle
\author Jouko Mickelsson \endauthor
\affil Department of Mathematics and Statistics, PL 68, FIN-00014 University of Helsinki, Finland; 
Department of Physics, Royal Institute of
Technology, SE-106 91 Stockholm, Sweden.\endaffil


\endtopmatter

Article in the Encyclopedia of Mathematical Physics, Elsevier Academic Press. Edited by
Jean-Pierre Francoise, Gregory L.  Naber, and Tsou Sheung Tsun.

\vskip 0.3in

\define\a{\alpha}

\redefine\b{\beta}
\define\g{\gamma}

\redefine\l{\lambda}

\define\<#1,#2>{\langle #1,#2\rangle}
\define\TR{\text{tr}}
\define\dep(#1,#2){\text{det}_{#1}#2}
\define\norm(#1,#2){\parallel #1\parallel_{#2}}

\document

\NoRunningHeads

1. DEFINITIONS AND AN EXAMPLE 

\vskip 0.3in

A gerbe can be viewed as a next step in a ladder of geometric and
topological
objects on a manifold which starts  from ordinary complex valued
functions and in the second step of sections of complex line bundles.

It is useful to recall the construction of complex line bundles and
their connections. Let $M$ be a smooth manifold and $\{U_{\alpha}\}$ an
of open cover  of $M$ which trivializes a line bundle $L$ over $M.$
Topologically, up to equivalence, the line bundle is completely 
determined by its Chern class which is a cohomology class $[c] \in 
H^2(M,\Bbb Z).$ On each open set $U_{\a}$ we may write $2\pi i c= dA_{\a},$ 
where $A_{\a}$ is a 1-form. On the overlaps $U_{\a\b}= U_{\a}\cap
U_{\b}$ we can write
$$ A_{\a} - A_{\b} = f_{\a\b}^{-1} df_{\a\b},\tag1$$
at least when $U_{\a\b}$ is contractible, where $f_{\a\b}$ is a circle 
valued complex function on the overlap. The data $\{c, A_{\a},
f_{\a\b}\}$ defines what is known as a (representative of a) Deligne 
cohomology class on the open cover $\{U_{\a}\}.$ The 1-forms $A_{\a}$ 
are the local potentials of the curvature form $2\pi ic$ and the
$f_{\a\b}$'s are the transition functions of the line bundle $L.$ 
Each of these three different data defines separately the equivalence 
class of the line bundle but together they define the line bundle with 
a connection. 

The essential thing is here that there is a bijection between the 
second integral cohomology of $M$ and the set of equivalence classes 
of complex line bundles over $M.$ It is natural to ask whether there 
is a geometric realization of integral third (or higher) cohomology. 
In fact, gerbes provide such a realization. Here we shall restrict 
to a smooth differential geometric approach which is by no means 
most general possible, but it is sufficient for most applications to 
quantum field theory. However, there are examples of gerbes over
orbifolds which do not need to come from finite group action on a
manifold, which are not covered by the following definition. 

For the examples in this article it is sufficient to adapt the
following definition. A gerbe over a manifold $M$ (without geometry)  
is simply a principal bundle $\pi: P\to M$ with fiber equal to
$PU(H),$ the projective unitary group of a Hilbert space $H.$ The
Hilbert space may be either finite or infinite dimensional.

The quantum field theory applications discussed in this article are related to
the \it chiral anomaly \rm for fermions in external fields. The link comes from
the fact that the chiral symmetry breaking leads in the generic case to 
\it projective representations \rm of the symmetry groups. For this reason, when 
modding out by the gauge or diffeomorphism symmetries, one is led to study 
bundles of projective Hilbert spaces. The anomaly is reflected as a nontrivial 
characteristic class of the projective bundle, known in mathematics literature 
as the \it Dixmier-Douady \rm class. 

In a suitable open cover the bundle $P$ has a family of local
trivializations with transition functions $g_{\a\b}: U_{\a\b} \to 
PU(H),$ with the usual cocycle property 
$$g_{\a\b} g_{\b\g} g_{\g\a}=1 \tag2 $$ 
on triple overlaps. Assuming that the overlaps are contractible, we
can chooce lifts $\hat g_{\a\b}: U_{\a\b} \to U(H),$ 
to the unitary group of the Hilbert space. However, 
$$\hat g_{\a\b} \hat g_{\b\g} \hat g_{\g\a} = f_{\a\b\g},\tag3$$ 
where the $f$'s are circle valued functions on triple overlaps.     
They satisfy automatically the cocycle property
$$f_{\a\b\g} f_{\a\b\delta}^{-1} f_{\a\g\delta} f_{\b\g\delta}^{-1}
=1 \tag4$$
on quadruple overlaps. There is an important difference between the 
finite and infinite dimensional cases. In the finite dimensional case 
the circle bundle $U(H) \to U(H)/S^1 = PU(H)$ reduces to a bundle 
with fiber $\Bbb Z/N\Bbb Z= \Bbb Z_N,$ where $N=\text{dim}\, H.$ 
This follows from $U(N)/S^1 = SU(N)/Z_N$ and the fact that $SU(N)$ is 
a subgroup of $U(N).$ For this reason one can choose the lifts $\hat
g_{\a\b}$ such that the functions $f_{\a\b\g}$ take values in the
  finite subgroup $\Bbb Z_N \subset S^1.$ 

The functions $f_{\a\b\g}$ define an element $a=\{a_{\a\b\g\delta}\}$ in the \v{C}ech cohomology 
$H^3(\Cal U, \Bbb Z)$ by a choice of logarithms, 
$$2\pi i a_{\a\b\g\delta} = \log f_{\a\b\g} -\log f_{\a\b\delta}+\log
f_{\a\g\delta} -\log f_{\b\g\delta}.\tag5$$  
In the finite dimensional case the \v{C}ech cocycle is necessarily
torsion, $Na=0,$ but not so if $H$ is infinite dimensional. In the
finite dimensional case (by passsing to a good cover and using the
\v{C}ech - de Rham equivalence over real or complex numbers) the class is 
third de Rham cohomology constructed from the transition functions 
is necessarily zero. Thus in general one has to work with \v{C}ech
cohomology to preserve torsion information. 
One can prove:

\proclaim{Theorem} The construction above is a one-to-one map between the set of equivalence
 classes of $PU(H)$ bundles over $M$  and  elements of $H^3(M,\Bbb Z).$ 
\endproclaim 

The characteristic class in $H^3(M,\Bbb Z)$ of a $PU(H)$ bundle is called the
\it Dixmier-Douady class. \rm 

\bf First example. \rm Let $M$ be an oriented Riemannian manifold and 
$FM$ its bundle of oriented orthonormal frames. The structure group of  
$FM$ is the rotation group $SO(n)$ with $n=\text{dim}\, M.$ The spin
bundle (when it exists) is a double covering Spin($M$) of $FM,$ with 
structure group $Spin(n),$ a double cover of $SO(n).$ Even when the 
spin bundle does not exist there is always the bundle Cl($M$) of
Clifford algebras over $M.$ The fiber at $x\in M$ is the Clifford
algebra defined by the metric $g_x,$ i.e., it is the complex $2^n$ 
dimensional algebra generated by the tangent vectors $v\in T_x(m)$
with the defining relations
$$ \g(u) \g(v) + \g(v) \g(u) = 2 g_x(u,v).$$ 
The Clifford algebra has a faithful representation in
$N=2^{[n/2]}$ dimensions ($[x]$ is the integral part of $x$) such that 
$$\g(a\cdot u) = S(a) \g(u) S(a)^{-1}$$ 
where $S$ is an unitary  representation of $Spin(n)$ in $\Bbb C^N.$ 
Since $Spin(n)$ is a double cover of $SO(n),$ the representation $S$  
may be viewed as a projective representation of $SO(n).$ Thus again,
if the overlaps $U_{\a\b}$ are contractible, we may choose a lift of
the frame bundle transition functions $g_{\a\b}$ to unitaries $\hat
g_{\a\b}$ in  $H= \Bbb C^N.$  In this case the functions $f_{\a\b\g}$ 
reduce to $\Bbb Z_2$ valued functions and the obstruction to the
lifting problem, which is the same as the obstruction to the existence 
of spin structure, is an element of $H^2(M,\Bbb Z_2),$ known as 
the second Stiefel-Whitney class $w_2.$ The image of $w_2$ with respect to
\it the Bockstein map \rm (in this case, given by the formula (5) )
gives a 2-torsion element in $H^3(M,\Bbb Z),$ the Dixmier-Douady 
class. 

Another way to think of a gerbe is the following (we shall see that this
arises in a natural way in quantum field theory).  There is a canonical 
complex line bundle $L$ over $PU(H),$ the associated line bundle to
the circle bundle $S^1 \to U(H) \to PU(H).$  Pulling back $L$ by the 
local transition functions $g_{\a\b} \to PU(H)$ we obtain a family of
line bundles $L_{\a\b}$  over the open sets $U_{\a\b}.$ By the cocycle property
(2) we have natural isomorphisms 
$$  L_{\a\b} \otimes  L_{\b\g} = L_{\a\g}.\tag6$$

We can take this as a definition of a gerbe over $M:$  A collection of
line bundles over intersections of open sets in an open cover of $M,$
satisfying the cocycle condition (6).   By (6) we have a trivialization
$$ L_{\a\b} \otimes L_{\b\g} \otimes L_{\g\a} = f_{\a\b\g} \cdot \bold{1},\tag7$$
where the $f$'s are circle valued functions on the triple overlaps.
By the Theorem  we conclude that indeed the data in (6) defines 
(an equivalence class of) a principal $PU(H)$ bundle.

If $L_{\a\b}$ and $L_{\a\b}'$ are two systems of local line bundles over the same 
cover, then the gerbes are equivalent if there is a system of line bundles $L_{\a}$ over 
open sets $U_{\a}$ such that 
$$L'_{\a\b} = L_{\a\b} \otimes L_{\a}^* \otimes L_{\b}\tag 8$$  
on each $U_{\a\b}.$ 

A gerbe  may come equipped with geometry, encoded in a Deligne 
cohomology class with respect to a given open covering of $M.$ 
The Deligne class is given by  functions $f_{\a\b\g},$ 1-forms 
$A_{\a\b},$ 2-forms $F_{\a},$ and a global 3-form (the Dixmier-Douady 
class of the gerbe) $\Omega,$ subject to the conditions
$$ \align dF_{\a}  & = 2\pi i \Omega \\
F_{\a} - F_{\b} &= dA_{\a\b} \\
A_{\a\b} - A_{\a\g} + A_{\b\g} &= f^{-1}_{\a\b\g}  df_{\a\b\g}. \tag9\endalign$$

\vskip 0.3in 
2. GERBES FROM CANONICAL QUANTIZATION 

\vskip 0.3in

Let $D_x$ be a family of self-adjoint Fredholm operators in a complex Hilbert space $H$ 
parametrized by $x\in M.$ This situation arises in quantum field theory for
example when $M$ is some space of external fields, coupled to Dirac 
operator $D$ on a compact manifold. The space $M$ might consists of 
gauge potentials (modulo gauge transformations) or $M$ might be the moduli space 
of Riemann metrics. In these examples the essential spectrum of $D_x$ is both 
positive and negative and the family $D_x$ defines an element of $K^1(M).$ In fact,
one of the definitions of $K^1(M)$ is that its elements are homotopy classes of
maps from $M$ to the space $\Cal F_*$ of self-adjoint Fredholm operators with both
positive and negative essential spectrum.  In physics applications one deals most
often with unbounded hamiltonians, and the operator norm topology    must be
replaced by something else; popular choises are the Riesz topology defined by
the map $F\mapsto  F/(|F| +1)$ to bounded operators or the gap topology defined 
by graph metric. 

The space $\Cal F_*$ is homotopy equivalent to the group $G=U_1(H)$ of unitary
operators $g$ in $H$ such that $g-1$ is a trace-class operator.  This space is 
a classifying space for principal $U_{res}$ bundles where $U_{res}$ is the group
of unitary operators $g$ in a \it polarized \rm complex  Hilbert space $\Cal H =
\Cal H_+ \oplus \Cal H_-$ such that the off-diagonal blocks of $g$ are Hilbert-Schmidt 
operators.  This is related to Bott periodicity. There is a natural principal bundle
$P$ over $G=U_1(H)$ with fiber equal to the group $\Omega G$ of based loops
in $G.$ The total space $P$  consists of smooth paths $f(t)$ in $G$ starting from the neutral
element such that  $f^{-1} df$ is smooth and periodic. The projection $P\to G$ is
the evaluation at the end point $f(1).$ The fiber is clearly $\Omega G.$ 
By Bott periodicity, the homotopy groups of $\Omega G$ are shifted from those of
$G$ by one dimension, i.e.,
$$\pi_{n} \Omega G = \pi_{n+1} G.$$ 
The latter are zero in even dimensions and equal to $\Bbb Z$ in odd dimensions.
On the other hand, it is known that the even homotopy groups of $U_{res}(\Cal H)$ are 
equal to $\Bbb Z$ and the odd ones vanish. In fact, with a little more effort 
one can show that the embedding of $\Omega G$ to $U_{res}(\Cal H)$ is a homotopy
equivalence,  when $\Cal H= L^2(S^1, H),$ the polarization being the splitting to
nonnegative and negative Fourier modes and the action  of $\Omega G$ is the pointwise
multiplication on $H$ valued functions on the circle $S^1.$ 

Since $P$ is contractible, it is indeed the classifying bundle for $U_{res}$ bundles.
Thus we conclude that $K^1(M)$ = the set of homotopy classes of maps $M\to G$ 
= the set of equivalence classes of $U_{res}$ bundles over $M.$  The relevance of this
fact in quantum field theory follows from the properties of representations of the \it
algebra of canonical anticommutation relations (CAR) \rm. For any complex Hilbert space
$H$ this algebra is the algebra generated by elements $a(v)$ and $a^*(v),$
with $v\in H$, subject to the relations
$$a^*(u) a(v) + a(v) a^*(u) = 2<v,u>,$$
where the Hilbert space inner product on the right is antilinear in the first argument, 
and all other anticommutators vanish. In addition, $a^*(u)$ is linear and $a(v)$ 
antilinear in its argument. 

An \it irreducible Dirac representation  \rm of the CAR algebra is given by a polarization
$H=H_+ \oplus H_-.$ The representation is characterized by the existence of a vacuum 
vector $\psi$ in the \it fermionic Fock space \rm $\Cal F$ such that 
$$ a^*(u) \psi=0 = a(v) \psi \text{ for } u\in H_-, \,\, v\in H_+. \tag10$$ 
A theorem of D. Shale and W.F. Stinespring says that two Dirac representations defined 
by a pair of polarizations $H_+,H_+'$ are equivalent if and only if
there is  $g\in U_{res}(H_+\oplus H_-)$ such that $H'_+ = g\cdot H_+.$ 
In addition, in order that a unitary tranformation $g$ is \it implementable \rm  in the
Fock space, i.e., there is a unitary operator $\hat g$ in $\Cal F$ such that 
$$ \hat g a^*(v) \hat g^{-1} = a^*(gv) \,,\,  \forall v\in H,\tag11$$
and similarly for the $a(v)$'s, one must have $g\in U_{res}$ with respect to
the polarization defining the vacuum vector. This condition is both necessary and
sufficient. 
 
 The polarization of the 1-particle Hilbert space comes normally from a spectral
 projection onto the positive energy subspace of a Hamilton operator. 
 In the background field problems one studies families of Hamilton operators 
 $D_x$ and then one would like to construct a family of fermionic Fock spaces
 parametrized by $x\in M.$ If none of the Hamilton operators has zero modes,
 this is unproblematic. However, the presence of zero modes makes it 
 impossible to define  the positive energy subspace $H_+(x)$ as a continuous 
 function of $x.$  One way out of this is to weaken the condition for the polarization:
 Each $x\in M$ defines a Grassmann manifold $Gr_{res}(x)$  consisting of all subspaces 
 $W\subset H$ such that the projections onto $W$ and $H_+(x)$ differ by 
 Hilbert-Schmidt operators. The definition of $Gr_{res}(x)$ is stable with respect to 
 finite rank perturbations of $D_x/|D_x|.$  For example, when $D_x$ is a Dirac operator on
 a compact manifold   then $(D_x -\l)/|D_x-\l |$ defines the same Grassmannian for
 all real numbers $\l$ because in each finite interval there are only a finite number 
 of eigenvalues (with multiplicities) of $D_x.$ From  this follows that the Grassmannians form 
 a locally trivial fiber bundle $Gr$  over families of Dirac operators. 
 
 If the bundle $Gr$ has a global section $x\mapsto  W_x$ then we can define a bundle of
 Fock space representations for the CAR algebra over the parameter space $M.$ 
 However, there are important situations when no global sections exist.  
 It is easier to explain the potential obstruction in terms of a principal
 $U_{res}$ bundle $P$ such that $Gr$ is an associated bundle to $P.$
 
 The fiber of $P$ at $x\in M$ is the set of all unitaries $g$ in $H$ such that 
 $g\cdot H_+ \in Gr_x$ where $H=H_+ \oplus H_-$ is a fixed reference polarization.
 Then we have 
 $$Gr =  P\times_{U_{res}} Gr_{res},$$
 where the right action of $U_{res}=U_{res}(H_+\oplus H_-)$ in the fibers of $P$ is the right multiplication
 on unitary operators and the left action on $Gr_{res}$ comes from the observation
 that $Gr_{res}= U_{res}/(U_+\times U_-)$ where $U_{\pm}$ are the diagonal 
 block matrices in $U_{res}.$ By a result of N. Kuiper, the subgroup $U_-\times U_-$ is
 contractible and so $Gr$ has a global section if and only if $P$ is trivial. 
 
 Thus in the case when $P$ is trivial we can define the family of Dirac representations 
of the CAR algebra parametrized by $M$ such that in each of the Fock spaces we have a Dirac 
vacuum which is in a precise sense close to the vacuum defined by the energy polarization.
However, the triviality of $P$ is not a necessary condition. Actually, what is needed is 
that $P$ ha a \it prolongation \rm to a bundle $\hat P$ with fiber $\hat U_{res}.$ The group
$\hat U_{res}$ is a central extension of $U_{res}$ by the group $S^1.$ 

The Lie algebra $\hat \bold{u}_{res}$ is as a vector space the direct sum $\bold{u}_{res}
\oplus i\Bbb R,$ with commutators
$$ [X + \l, Y+ \mu] = [X,Y] + c(X,Y),\tag12$$ 
where $c$ is the Lie algebra cocycle 
$$c(X,Y)= \frac14 \TR\, \epsilon[\epsilon,X][\epsilon,Y].\tag13$$

Here $\epsilon$ is the grading operator with eigenvalues $\pm 1$ on $H_{\pm}.$ 
The trace exists since the off diagonal blocks of $X,Y$ are Hilbert-Schmidt. 

The group $\hat U_{res}$ is a circle bundle over $U_{res}.$ The Chern class of the associated 
complex line bundle is the generator of $H^2(U_{res}, \Bbb Z)$ and is given explicitely at the 
identity element as the antisymmetric bilinear form $c/2\pi i$ and at other points on the group manifold
through left-translation of $c/2\pi i.$ If $P$ is trivial, then it has an obvious prolongation to the
trivial bundle $M\times \hat U_{res}.$ In any case, if the prolongation exists we can define the 
bundle of Fock spaces carrying CAR representations as the associated bundle 
$$\Cal F= \hat P \times_{\hat U_{res}} \Cal F_0,$$ 
where is $\Cal F_0$ is the fixed Fock space defined by the same polarization $H=H_+\oplus H_-$ 
used to define $U_{res}.$ By the Shale-Stinespring theorem, 
any $g\in U_{res}$ has an implementation 
$\hat g$ in $\Cal F_0,$ but $\hat g$ is only defined up to phase, thus the central $S^1$ extension. 

The action of the CAR algebra in the fibers is given as follows. For $x\in M$ choose any $\hat g\in \hat P_x.$ 
Define 
$$ a^*(v) \cdot (\hat g, \psi) = (\hat g, a^*(g^{-1}v) \psi),$$ 
where $\psi\in \Cal F_0$ and $v\in H;$ similarly for the operators $a(v).$ It is easy to check that 
this definition passes to the equivalence classes in $\Cal F.$ Note that the representations in different 
fibers are in general inequivalent because the tranformation $g$ is not implementable in the Fock space
$\Cal F_0.$ 

The potential obstruction to the existence of the prolongation of $P$ is again a 3-cohomology class on the 
base. Choose a good cover of $M.$ On the intersections $U_{\a\b}$ of the open cover the transition functions
$g_{\a\b}$ of $P$ can be prolonged to functions $\hat g_{\a\b}: U_{\a\b} \to \hat U_{res}.$ We have 
$$\hat g_{\a\b} \hat g_{\b\g} \hat g_{\g\a} = f_{\a\b\g} \cdot 1,\tag14$$ 
for functions $f_{\a\b\g}: U_{\a\b\g} \to S^1,$ which by construction satisfy the 
cocycle property (4). Since the cocycle is defined on a good cover, it defines an integral
\v{C}ech cohomology class $\omega \in H^3(M,\Bbb Z).$

Let us return to the universal $U_{res}$ bundle $P$ over $G=U_1(H).$ In this case the prolongation 
obstruction can be computed relativly easily. It turns out that the 3-cohomology class is represented 
by the de Rham class which is the generator of $H^3(G,\Bbb Z).$ Explicitly,
$$ \omega = \frac{1}{24 \pi^2} \TR\, (g^{-1} dg)^3.\tag15 $$ 
Any principal $U_{res}$ bundle over $M$ comes from a pull-back of $P$ with respect to a map 
$f:M\to G,$ so the Dixmier-Douady class in the general case is the pull-back $f^*\omega.$ 

The line bundle construction of the gerbe over the parameter space $M$ for Dirac operators is given by the 
observation that the spectral subspaces $E_{\l\l'}(x)$ of $D_x,$ corresponding to the open interval 
$]\l,\l'[$ in the real line, form finite rank vector bundles over open sets $U_{\l\l'}=U_{\l} \cap 
U_{\l'}.$ Here $U_{\l}$ is the set of points $x\in M$ such that $\l$ does not belong to the spectrum
of $D_x.$ Then we can define, as top exterior power, 
$$L_{\l\l'} = {\bigwedge}^{top}(E_{\l\l'})$$ 
as the complex vector bundle over $U_{\l\l'}.$ It follows immediately from the definition that the 
cocycle property (6) is satisfied. 

\bf Example 1.  Fermions on an interval. \rm Let $K$ be a compact group and $\rho$ its unitary 
repsentation in a finite-dimensional vector space $V.$ Let $H$ be the Hilbert space of square-integrable 
$V$ valued functions on the interval $[0,2\pi]$ of the real axis. For each $g\in K$ let $Dom_g\subset H$ be 
the dense subspace of smooth functions $\psi$ with the boundary condition $\psi(2\pi)= \rho(g)\psi(0).$ 
Denote by $D_g$ the operator $-i\frac{d}{dx}$ on this domain. The spectrum of $D_g$ is a function of the 
eigenvalues $\l_k$ of $\rho(g),$ consisting of real numbers  $n+\frac{\log(\l_k)}{2\pi i} $ with $n\in \Bbb Z.$
For this reason the splitting of the 1-particle space $H$ to positive and negative modes of the operator
$D_g$ is in general not continuous as function of the parameter $g.$ This leads to the problems described
above. However, the principal $U_{res}$ bundle  can be explicitly constructed. It is the pull-back of the 
universal bundle $P$ with respect to the map $f:K \to G$ defined by the embedding $\rho(K) \subset G$ as 
$N\times N$ block matrices, $N= \text{dim}\, V.$ Thus the Dixmier-Douady class in this example is
$$\omega= \frac{1}{24\pi^2} \TR\, (\rho(g)^{-1} d \rho(g))^3.\tag16$$ 

\bf Example 2. Fermions on a circle.  \rm Let $H= L^2(S^1,V)$ and $D_A= -i(\frac{d}{dx} + A)$ where 
$A$ is a smooth vector potential on the circle taking values in the Lie algebra $\bold{k}$ of $K.$ 
In this case the domain is fixed, consisting of smooth $V$ valued functions on the circle. The $\bold{k}$ 
valued function $A$ is represented as a multiplication operator through the representation $\rho$ of $K.$ 
The parameter space $\Cal A$ of smooth vector potentials is flat, thus there cannot be any obstruction to 
the prolongation problem. However, in quantum field theory one wants to pass to the moduli space $\Cal A/\Cal G$ 
of gauge potentials. Here $\Cal G$ is the group of smooth based gauge transformations, i.e., $\Cal G= \Omega K.$ 
Now the moduli space is the group of holonomies around the circle, $\Cal A/\Cal G= K.$ Thus we are in a similar 
situation as in Example 1. In fact, these examples are really two different realizations of the same family of 
self-adjoint Fredholm operators. The operator $D_A$ with $k=holonomy(A)$ has exactly the same spectrum as $D_k$ 
in Example 1. For this reason the Dixmier-Douady class on $K$ is the same as before. 

The case of Dirac operators on the circle is simple because all the energy polarizations for different 
vector potentials are elements in a single Hilbert-Schmidt Grassmannian $Gr(H_+\oplus H_-)$ where we can 
take as the reference polarization the splitting to positive and negative Fourier modes. Using this polarization,
the bundle of fermionic Fock spaces over $\Cal A$ can be trivialized as $\Cal F= \Cal A\times \Cal F_0.$ 
However, the action of the gauge group $\Cal G$ on $\Cal F$ acquires a central extension $\hat \Cal G\subset 
\widehat{LK},$ where $LK$ is the free loop group of $K.$
The Lie algebra cocycle determining the central extension is 
$$ c(X,Y)= \frac{1}{2\pi i} \int_{S^1} \TR_{\rho} \,X d Y,\tag17$$ 
where $\TR_{\rho}$ is the trace in the representation $\rho$ of $K.$   Because of the central extension, the 
quotient $\Cal F/\hat \Cal G$ defines only a projective vector bundle over $\Cal A/\Cal G,$ the Dixmier-Douady 
class being given by (16).

In the Example 1 (and 2) above the complex line bundles can be constructed quite explicitly. Let us study the case
$K=SU(n).$ Define $U_{\l} \subset K$ as the set of matrices $g$ such that $\l$ is not an eigenvalue of $g.$ 
Select $n$ different points $\l_j$ on the unit circle such that their product is not equal to $1.$ We assume that 
the points are ordered counter clock-wise on the circle. Then the 
sets $U_j= U_{\l_j}$ form an open cover of $SU(n).$ On each $U_j$ we can choose a continuous branch of the logarithmic 
function $\log: U_j \to \bold{su(n)}.$ The spectrum of the Dirac operator $D_g$ with the holonomy $g$ consists 
of the infinite set of numbers $\Bbb Z + Spec(-i\log(g)).$ In particular, the numbers $\Bbb Z -i \log \l_j$ do not 
belong to the spectrum of $D_g.$  Choosing  $\mu_k= -i\log \l_k$ as an increasing sequence in the 
interval $[0,2\pi[$ we can as well define 
$U_j=\{x\in M| \mu_j \notin Spec(D_x)\}.$  In any case, the top exterior power of the spectral subspace $E_{\mu_j, \mu_k}(x)$ 
is given by zero Fourier modes consisting of the spectral subspace of the holonomy $g$ in the segment $]\l_j,\l_k[$ 
of the unit circle.

\vskip 0.3in

3. INDEX THEORY AND GERBES  

\vskip 0.3in

Gauge and gravitational anomalies in quantum field theory can be computed by Atiyah-Singer index theory.
The basic setup is as follows. On a compact even dimensional spin manifold $S$ (without boundary) the Dirac 
operators coupled to vector potentials and metrics form a family of Fredholm operators. The parameter space 
is the set $\Cal A$ of smooth vector potentials (gauge connections) in a vector bundle over $S$ and the 
set of smooth Riemann metrics on $S.$ The family of Dirac operators is covariant with respect to gauge
transformations and diffeomorphims of $S,$  thus we may view the Dirac operators parametrized by the 
moduli space $\Cal A/\Cal G$  of gauge connections and the moduli space $\Cal M/\text{Diff}_0(S)$ of Riemann 
mertics. Again, in order that the moduli spaces are smooth manifolds one has to restrict to the based 
gauge transformations, i.e., those which are equal to the neutral element in a fixed base point in each 
connected component of $S.$ Similarly, the Jacobian of a diffeomorphism is required to be equal to the 
identity matrix at the base points. Passing to the quotient modulo gauge transformations and diffeomorphims
we obtain a vector bundle over the space 
$$S\times \Cal A/\Cal G \times \Cal M/\text{Diff}_0(S).\tag18$$ 
Actually, we could as well consider a generalization in which the base space is a fibering over the moduli space 
with model fiber equal to $S,$ but for simplicity we stick to (18). 

According to Atiyah-Singer index formula for families, the K-theory class of the family of Dirac operators 
acting on the smooth sections of the tensor product of the spin bundle and the vector bundle $V$ over 
(18) is given through the differential forms
$$\hat A(R) \wedge ch(V),$$ 
where $\hat A(R)$ is the A-roof genus, a function of the Riemann curvature tensor $R$ associated to the
Riemann metric, 
$$\hat A(R) = \text{det}^{1/2}\left(  \frac{R/4\pi i}{\sinh(R/4\pi i)}\right),$$ 
and $ch(V)$ is the Chern character 
$$ch(V) = \TR\, e^{F/2\pi i},$$ 
where $F$ is the curvature tensor of a gauge connection. Here both $R$ and $F$ are forms on the infinite-dimensional 
base space (18). After integrating over the fiber $S,$ 
$$ Ind = \int_S \hat A(R) \wedge ch(V),\tag19$$ 
we obtain a family of differential forms $\phi_{2k},$  one in each even dimension, on the moduli space. 

The (cohomology classes of) forms $\phi_{2k}$ contain important topological information for the quantized 
Yang-Mills theory and for quantum gravity. The form $\phi_2$ describes potential \it chiral anomalies. \rm 
The chiral anomaly is a manifestation of gauge or reparametrization symmetry breaking. If the class 
$[\phi_2]$ is nonzero, the quantum effective action cannot be viewed as a function on the moduli space. Instead,
it becomes a section of a complex line bundle $DET$ over the moduli space. 

Since the Dirac operators are Fredholm (on compact manifolds), at a given point in the moduli space we can 
define the complex line 
$$DET_x =  {\bigwedge}^{top}(ker D_x^+) \otimes {\bigwedge}^{top}(coker D_x^+)\tag20$$ 
for the \it chiral \rm Dirac operators $D^+_x.$ In the even dimensional case the spin bundle is $\Bbb Z_2$ graded such that
the grading operator $\Gamma$ anticommutes with $D_x.$ Then $D_x^+ = P_- D_x P_+$ where $P_{\pm}= \frac12 
(1\pm \Gamma)$ are the chiral projections. $\wedge^{top}$ means the operation on finite dimensional vector 
spaces $W$ taking the exterior power of $W$ to dim$\, W.$ 

When the dimensions of the kernel and cokernel of $D_x$ are constant the formula (20) defines a smooth complex 
line bundle over the moduli space. In the case of varying dimensions a little extra work is needed 
to define the smooth structure.

The form $\phi_2$ is the Chern class of $DET.$ So if $DET$ is nontrivial, gauge covariant quantization of the 
family of Dirac operators is not possible. 

One can also give a geometric and topological meaning to the chiral symmetry breaking in Hamiltonian 
quantization and this leads us back to gerbes on the moduli space. Here we have to use an odd version 
of the index formula (19). Assuming that the physical space-time is even dimensional, at a fixed time 
the space is an odd dimensional manifold $S$. We still assume that $S$ is compact. In this case the integration 
in (19) is over odd dimensional fibers and therefore the formula produces a sequence of odd forms 
on the moduli space. 

The first of the odd forms $\phi_1$ gives the spectral flow of a 1-parameter family of operators $D_{x(s)}.$ Its 
integral along the path $x(t),$ after a correction by the difference of the eta invariant at the end points of the path,
in the moduli space, gives twice the difference of positive eigenvalues crossing over 
to the negative side of the spectrum minus the flow of eigenvalues in the opposite direction. The second term $\phi_3$ 
is the Dixmier-Douady class of the projective bundle of Fock spaces over the moduli space. 
In the Examples 1 and 2 the index theory calculation gives exactly the form (16) on $K.$ 

\bf Example \rm Consider Dirac operators on the 3-dimensional sphere $S^3$  coupled to vector potentials. Any vector 
bundle on $S^3$ is trivial, so let $V=S^3 \times \Bbb C^N.$ Take $SU(N)$ as the gauge group and let $\Cal A$ be the 
space of 1-forms on $S^3$ taking values in the Lie algebra $\bold{su}(N)$ of $SU(N).$ Fix a point $x_s$ on $S^3$, the 
'South Pole', and let $\Cal G$ be the group of gauge transformations based at $x_s.$ That is, $\Cal G$ consists of
smooth functions $g:S^3\to SU(N)$ with $g(x_s)=1.$ In this case $\Cal A/\Cal G$ can be identified as $Map(S^2, SU(N))$ 
times a contractible space. This is because any point $x$ on the equator of $S^3$ determines a unique semicircle from
the South Pole to the North Pole through $x.$  The parallel transport along this path with respect to a vector potential
$A\in\Cal A$ defines an element $g'_A(x) \in SU(N),$  using the fixed trivialization of $V.$ Set $g_A(x) = g'_A(x)g'_A(x_0)^
{-1}$ where $x_0$ is a fixed point on the equator. The element $g_A(x)$ then depends only on the gauge equivalence class 
$[A]\in \Cal A/\Cal G.$ It is not difficult to show that the map $A\mapsto g_A$ is a homotopy equivalence from the moduli 
space of gauge potentials to the group $\Cal G_2=Map_{x_0}(S^2, SU(N)),$ based at $x_0.$ When $N>2$ the cohomology $H^5(SU(N),\Bbb Z) 
=\Bbb Z$ transgresses to the cohomology $H^3(\Cal G_2, \Bbb Z) = \Bbb Z.$ In particular, the generator 
$$\omega_5 = \left(\frac{i}{2\pi}\right)^3 \frac{2}{5 !} \TR\, (g^{-1} dg)^5$$ 
of $H^5(SU(N),\Bbb Z)$ gives the generator of $H^3(\Cal G_2,\Bbb Z)$ by contraction and integration, 
$$\Omega = \int_{S^2} \omega_5.$$ 

\vskip 0.3in 

4. GAUGE GROUP EXTENSIONS 

\vskip 0.2in

The new feature for gerbes associated to Dirac operators in higher than one dimensions is that the gauge group, acting on the
bundle of Fock spaces parametrized by vector potentials, is represented through \it an abelian extension. \rm On the Lie algebra 
level this means that the Lie algebra extension is not given by a scalar cocycle $c$ as in the one dimensional case but by 
a cocycle taking values in an abelian Lie algebra. In the case of Dirac operators coupled to vector potentials the abelian
Lie algebra consists of a certain class of complex functions on $\Cal A.$ The extension is then defined by the commutators
$$[(X,\a), (Y,\b)] = ([X,Y], \Cal L_X \b -\Cal L_Y \a + c(X,Y)) \tag21$$ 
where $\a,\b$ are functions on $\Cal A$ and $\Cal L_X \b$ denotes the Lie derivative of $\b$ in the direction of the 
infinitesimal gauge transformation $X.$ The 2-cocycle property of $c$ is expressed as 
$$c([X,Y], Z) + \Cal L_X c(Y,Z) + \text{ cyclic permutations of $X,Y,Z$} = 0. $$ 
In the case of Dirac operators on a 3-manifold $S$ the form $c$ is the Mickelsson-Faddeev cocycle
$$c(X,Y) = \frac{i}{12\pi^2}  \int_S \TR_{\rho} \, A\wedge(dX\wedge dY -dY\wedge dX).\tag22$$ 

The corresponding gauge group extension is an extension of $Map(S,G)$ by the normal subgroup $Map(\Cal A, S^1).$ As 
a topological space, the extension is the product 
$$Map(\Cal A, S^1) \times_{S^1} P,$$ 
where $P$ is a principal $S^1$ bundle over $Map(S,G).$   

The Chern class $c_1$ of the bundle $P$ is again computed by transgression from $\omega_5,$ this time 
$$c_1 =  \int_{S} \omega_5  .$$ 
In fact, we can think the cocycle $c$ as a 2-form on the space of flat vector potentials $A=g^{-1} dg$ with
$g\in Map(S^3,G).$  Then one can show that the cohomology classes $[c]$ and $[c_1]$ are equal.

As we have seen, the central extension of a loop group is the key to understanding the quantum field theory gerbe.
Here is a short description of it starting from the 3-form (16) on a compact Lie group $G.$  First define a central 
extension $Map(D,G) \times S^1$ of the group of smooth maps from the unit disk $D$ to $G,$ with point-wise 
multiplication. The group multiplication is given as
$$ (g,\l)\cdot (g',\l')= (gg', \l\l' \cdot e^{2\pi i\gamma(g,g')}),$$ 
where 
$$\g(g,g') = \frac{1}{8\pi^2} \int_D \TR_{\rho} \, g^{-1}dg \wedge dg' {g'}^{-1},\tag23$$ 
where the trace is computed in a fixed unitary representation $\rho$ of $G.$  This group contains as a normal subgroup
the group $N$ consisting of pairs $(g, e^{2\pi i C(g)})$ with
$$C(g) = \frac{1}{24\pi^2} \int_B \TR_{\rho} \, (g^{-1} dg)^3.\tag24$$ 
Here $g(x)=1$ on the boundary circle $S^1=\partial D,$ and thus can be viewed as a function $S^2 \to G.$ The 3-dimensional
unit ball $B$ has $S^2$ as a boundary and $g$ is extended in an arbitrary way from the boundary to the ball $B.$ 
The extension is possible since $\pi_2(G) =0$ for any finite-dimensional Lie group. The value of $C(g)$ depends on the extension  
only modulo an integer and therefore $e^{2\pi iC(g)}$ is well-defined.

The central extension is then defined as 
$$ \widehat{LG} = (Map(D,G) \times S^1)/N.$$ 
One can show easily that the Lie algebra of $\widehat{LG}$ is indeed given through the cocycle (17).  In the case of $G=SU(n)$ in the
defining representation, this central extension is the basic extension: The cohomology class is the generator of $H^2(LG,\Bbb Z).$ 
In general, to obtain  the basic extension one has to correct (23)  and (24) by a normalization factor.

This construction generalizes to the higher loop groups $Map(S,G)$ for compact odd dimensional manifolds $S.$ 
For example, in the case of a 3-manifold one starts from an extension of $Map(D,G),$ where $D$ is a 4-manifold 
with boundary $S.$ The extension is defined by a 2-cocycle $\gamma,$
but now for given $g,g'$ the cocycle $\g$ is a real valued  function 
of a point $g_0\in Map(S,G),$  which is  a certain  differential polynomial in the Maurer-Cartan 1-forms 
$g_0^{-1} dg_0,g^{-1}dg, g^{-1}dg.$  The normal subgroup $N$ is defined in a similar way, now $C(g)$ is the integral of 
the 5-form $\omega_5$ over a 5-manifold $B$ with boundary $\partial B$ identified as $D/\sim,$ the equivalence 
shrinking the boundary of $D$ to one point.  This gives the extension only over the conneted component of identity in
$Map(S,G),$ but it can be generalized to the whole group. For example, when $S=S^3$  and $G$ is simple the
connected components are labeled by elements of the third homotopy group $\pi_3 G =\Bbb Z.$ 

In some cases the de Rham cohomology class of the extension vanishes but the extension still contains 
interesting torsion information. In quantum field theory this comes from hamiltonian formulation of \it global 
anomalies. \rm A typical example of this phenomen is the Witten $SU(2)$ anomaly in four space-time dimensions.
In the hamiltonian formulation we take $S^3$ as the physical space, the gauge group $G=SU(2).$ In this case
the second cohomology of $Map(S^3,G)$ becomes pure torsion, related to the fact that the 5-form $\omega_5$ 
on $SU(2)$ vanishes for dimensional reasons. Here the homotopy group $\pi_4(G)=\Bbb Z_2$ leads the the nontrivial 
fundamental group $\Bbb Z_2$ in each connected component of $Map(S^3, G).$ Using this fact one can show that there 
is a nontrivial $\Bbb Z_2$ extension of the group $Map(S^3,G).$

\vskip 0.3in 
\bf Related material in this Encyclopedia:  \rm Articles  149 (Index theorems), 287 (Anomalies),
305 (Dirac fields and Dirac operators), 308 (Bosons and fermions in external fields),
322 (K-theory), 354 (Characteristic classes). 

\vskip 0.3in

\bf Further reading  \rm 

\vskip 0.3in

H. Araki: Bogoliubov automorphisms and Fock representations of
canonical anticommutation relations   In: Contemporary Mathematics, vol.
\bf 62, \rm American Mathematical Society, Providence, 1987.

N. Berline, E. Getzler, and M. Vergne: \it Heat Kernels and Dirac operators. \rm Die Grundlehren der mathematischen
Wissenschaften \bf 298, \rm  Springer Verlag, Berlin, 1992.

J.-L. Brylinski: \it  Loop Spaces, Characteristic Classes, and Geometric Quantization. \rm
Progress in Mathematics \bf 107, \rm Birkh\"auser Boston Inc., Boston, MA, 1993 

A.L. Carey, J. Mickelsson, and M. Murray: Bundle gerbes applied to quantum field theory. 
Rev. Math. Phys. \bf 12, \rm 65-90, 2000 

K. Gawedzki and N. Reis: WZW branes and gerbes. Rev.Math.Phys.\bf 14, \rm 1281-1334, 2002

J. Giraud: \it Cohomologie non abelienne. \rm (French)  Die Grundlehren der Mathematischen Wissenschaften \bf 179. \rm 
Springer Verlag, Berlin, 1971 

N. Hitchin: Lectures on special Lagrangian submanifolds. \it  Winter School on Mirror Symmetry,
 Vector Bundles and Lagrangian Submanifolds (Cambridge, MA, 1999),  \rm 151--182. In:  AMS/IP Stud. Adv. Math., \bf 23,  \rm  Amer. Math. Soc., Providence, RI, 2001.

J. Mickelsson: \it Current Algebras and Groups. \rm Plenum Press, London, 1989

S.B  Treiman, R.  Jackiw, B.  Zumino, and E. Witten: \it  Current algebra and anomalies. \rm Princeton Series in Physics.
 Princeton University Press, Princeton, NJ, 1985.

\enddocument